\documentclass[12pt]{article}

\usepackage[numbers, square, compress]{natbib}

\usepackage{epsf}
\usepackage{graphicx}
\usepackage{psfrag}
\usepackage[dvips]{epsfig}

\usepackage{amsmath}

\usepackage{verbatim}

\usepackage[cp866]{inputenc}
\usepackage[english]{babel}
\usepackage{amssymb,indentfirst}
\usepackage{hyperref}

\makeatletter
\renewcommand \thesection {\@arabic\c@section.}
\renewcommand\thesubsection   {\thesection\@arabic\c@subsection.}
\renewcommand\thesubsubsection{\thesubsection\@arabic\c@subsubsection.}
\makeatother

\def\lineup#1{\mbox{$\raise1.0ex\hbox{--} \kern-0.6em#1$}}

\def\starup#1{\mbox{$\raise1.6ex\hbox{$*$} \kern-0.5em#1$}}
\def\starupp#1{\mbox{$\raise1.8ex\hbox{$*$} \kern-1.0em#1$}}
\def\staruppp#1{\mbox{$\raise1.0ex\hbox{$*$} \kern-0.5em#1$}}
\def\sstarup#1{\mbox{\scriptsize $\raise1.8ex\hbox{$*$} \kern-.7em#1$}}

\def\ttildeup#1{\mbox{$\raise0.0ex\hbox{\Large $\; \tilde{}$} \kern-0.45em#1$}}
\def\bbar#1{\mbox{$\raise-0.4ex\hbox{\Large $\; \bar{}$} \kern-0.35em#1$}}

\begin{document}

\title{
  Chiral gauge leptoquark mass limits and branching ratios of
$ K_L^0, B^0, B_s \to l^+_i l^-_j  $ decays with account
 of the general fermion mixing in leptoquark currents
 }

\author{ M.~V.~Martynov\footnote{E-mail: martmix@mail.ru},
A.~D.~Smirnov\footnote{E-mail: asmirnov@uniyar.ac.ru} \\
{\small\it Division of Theoretical Physics, Department of Physics,}\\
{\small\it Yaroslavl State University, Sovietskaya 14,}\\
{\small\it 150000 Yaroslavl, Russia.}}
\date{}
\maketitle

\begin{abstract}
\noindent
The contributions of the chiral gauge leptoquarks  $V^{L,R}$
induced by the chiral four color quark-lepton symmetry
to the branching ratios of $ K_L^0, B^0, B_s \to l_1 \, l_2 $ decays
are calculated and analysed using the general parametrizations
of the fermion mixing matrices in the leptoguark currents.
From the current experimental data on these decays
under assumption $ m_{V^L} \ll m_{V^R} $
the lower mass limit $ m_{V^L} \cos{\gamma_L} > 5.68  \,\, \mbox{TeV}$
is found, which in particular case of equal gauge coupling constants
gives $ m_{V^L} > 8.03 \,\, \mbox{TeV} $.
The branching ratios of the decays under consideration
predicted by the chiral gauge leptoquarks are calculated and analysed
in dependence on the leptoquark masses and the mixing
parameters. It is shown that in consistency with the current experimental data
these branching ratios for $ B_s, B^0 \to \mu e $ decays can be close
to their experimental limits and those for $ B_s, B^0  \to  \tau e, \tau \mu  $
decays can be of order of~$10^{-7}$.
The calculated branching ratios will be useful in
the further experimental searches for these decays.


%
\vspace{3mm}

\vspace{3mm}

\end{abstract}




\vspace{3mm}


The search for a new physics beyond the Standard Model (SM) is one
of the directions of the modern  
studies in the 
high energy physics.
There is a lot of possible variants of new physics (such as supersymmetry,
left-right symmetry, two Higgs model, extended dimension models, etc.)
which are now under theoretical discussions and experimental searches
(including the experimental searches at the LHC).

One such variant of new physics can be induced by
the possible four color symmetry between quarks and leptons regarding leptons
as the quarks of the fourth color.
Proposed firstly \cite{Pati:1974yy} on the basis of the gauge group
 $G_{PS} = SU_V(4)\times SU_L(2)\times SU_R(2)$
this symmetry can be unified with the SM 
in the minimal way in frame of the 
model 
based on the gauge group \cite{Smirnov:1995jq,AD22,Perez:2013osa}
 $G_{MQLS}=SU_V(4) \times SU_L(2) \times U_R(1)$, where the last two factors
correspond to the usual electroweak symmetry of the SM.
These groups can be embedded into
the GUT group $SO(10)$~\cite{Fritzsch:1974nn,Chanowitz:1977ye,Georgi:1979dq}
as the intermediate stages in appropriate schemes~\cite{Ferrari:2018rey}
of the $SO(10)$ symmetry breaking.
In both cases the four color symmetry is described by the group $SU_V(4)$ of
the vector-like type and predicts in particular 
the new gauge particles --
the vector leptoquarks   
which belong to the 15-plet
 of the group $SU_V(4)$ and form the color triplet $(3, 1)_{2/3}$
of the SM group $G_{SM}~=~SU_c(3)~\times~SU_L(2)~\times~U(1)$.

The vector-like group $SU_V(4)$ can be extended to the gauge group of
the chiral four color symmetry for example in frame of group \cite{Fornal:2018dqn}
$SU_L(4) \times SU_R(4) \times SU_L(2) \times U(1)^{\prime}$.
In this case the chiral four color symmetry predicts two types of
the chiral gauge leptoquarks which are the color triplets $(3, 1)_{2/3}$
of the SM group 
and each of them interacts with left-handed and right-handed leptoquark
currents separately.
The possibility of the chiral four color symmetry has been also considered
in Refs. \cite{Smirnov:2007hv, Smirnov:2008zzb} in context of the mass limits
for the leptoquarks.
In general case the leptoquarks are the gauge or scalar particles carrying
both the  baryon and lepton numbers. They appear in many models and can led
to varied new physics effects, the comprehensive review of the physics of leptoquarks
can be found in Ref.~\cite{Dorsner:2016wpm}.

In the last years the leptoquarks are intensively used for the possible
explanations of the known anomalies in the semileptonic $B$ meson decays.
In this case to obviate the existing high limits on the masses of
the vector leptoquarks the conventional Pati-Salam four color symmetry
is subjected to modifications by the appropriate choice of 
couplings~\cite{Sahoo:2016pet,Sahoo:2018ffv} or mixings~\cite{Assad:2017iib} in fermion sector,
by introducing the additional factor $SU(3)^{\prime}$~\cite{DiLuzio:2017vat,DiLuzio:2018zxy}
with the third family quark-lepton unification~\cite{Greljo:2018tuh},
by PS triplication~\cite{Bordone:2017bld},
by using also the scalar leptoquarks~\cite{Heeck:2018ntp,Faber:2018qon,Faber:2018afz,Malinsky:2019tck},
by extending the fermion sector of the Pati-Salam model~\cite{Calibbi:2017qbu}
with introducing the nonunitary mixings due to additional vector-like
heavy leptons~\cite{Hati:2019ufv}.

The mass limits for the chiral gauge leptoquarks are more mild and
by this reason such leptoquarks look as the more natural ones for explanations
of the $B$ anomalies.
The gauge leptoquarks which couple to quarks and leptons in a predominately
left-handed manner were considered in the model with the gauge symmetry
$ SU(4)_C \times SU(2)_L \times U(1)_{Y^\prime}$~\cite{Balaji:2018zna}
\footnote{ A possibility to lower the mass scale of
the $SU_C(4)\times SU_L(2)\times SU_R(2)$ symmetry by the leptoquark interaction
only with the right-handed currents and introducing additional exotic leptons
was considered some time ago in Refs.~\cite{Foot:1997pb,RF2}. }
and with accounting also the new scalar leptoquarks in frame of the Pati-Salam-like
group $SU_C(4)\times SU_L(2)\times SU_R(2)$ with unification of the left-handed quarks
and leptons into a fundamental representation of $SU_C(4)$ and with a separate treatment
of right-handed quarks and leptons~\cite{Balaji:2019kwe}.
The chiral leptoquarks as the gauge bosons of the chiral four color symmetry
were considered in the model with the gauge symmetry
$SU_L(4) \times SU_R(4) \times SU_L(2) \times U(1)^{\prime}$~\cite{Fornal:2018dqn}
with the subgroup $SU_R(4)$ assumed to be broken at a much higher scale than $SU_L(4)$,
leading to a suppression of right-handed leptoquark currents.

The possible effects of leptoquarks in experiments depend on the masses of leptoquarks.
The lower mass limits for leptoquarks from their direct searches are
of about or less $1 \,\, \mbox{TeV}$. The essentially more stringent lower mass limits
are resulting from the rare decays of pseudoscalar mesons of the type
\begin{eqnarray}
K_L^0, B^0, B_s \to l^+_i l^-_j .
\label{eq:KL0B0Bsdecays}
\end{eqnarray}
The most stringent of them are resulting from the $K^0_L \to e^{\mp} \mu^{\pm}$ decay
and with neglect of fermion mixing in leptoquark currents are of order of
$2\,000 \,\, \mbox{TeV}$ for the vector
leptoquark~\cite{Valencia:1994cj,KM1,Kuznetsov:1995wb,Smirnov:2007hv,Smirnov:2008zzb}
and of order of $260 \,\, \mbox{TeV}$ for the chiral one~\cite{Smirnov:2007hv,Smirnov:2008zzb}.
These mass limits can be essentially lowered by account of fermion mixing in leptoquark
currents~\cite{Kuznetsov:2012ai,Smirnov:2017noz,Smirnov:2018ske,Povarov:2019fcz}
and instead of $2\,000 \,\, \mbox{TeV}$ the current lower mass limit for the vector
leptoquark with account of the fermion mixing in the leptoguark currents
of the general form is of order of $90 \,\, \mbox{TeV}$~\cite{Smirnov:2018ske}.
It is interesting now to know what can be the lower mass limits for the chiral
gauge leptoquarks resulting from the decays~(\ref{eq:KL0B0Bsdecays})
with account of the fermion mixing in the leptoguark currents.

In this paper the new lower mass limit for the chiral gauge leptoquarks
which results from the current experimental data on the decays~(\ref{eq:KL0B0Bsdecays})
is obtained with account of the general parametrizations
of the fermion mixing matrices in the leptoquark currents.
The contributions of the chiral gauge leptoquarks
to the branching ratios of these decays are calculated
and analysed in dependence on the leptoquark masses and mixing angles and phases
in comparision with the corresponding current data on these decays.
The leptoquarks under consideration are regarded as the gauge bosons
of the chiral four color symmetry group
\begin{eqnarray}
G_c = SU_c^L(4) \times SU_c^R(4),
\label{eq:Gc4L4R}
\end{eqnarray}
where the left( right )-handed quarks and leptons of each generation
(after fermion mixing) are unified into a fundamental representation
of $SU_c^L(4)$  ( $SU_c^R(4)$ ) group.



The interaction of the chiral gauge leptoquarks $V^L_{\alpha}, \, V^R_{\alpha}$
with down quarks $d_{p \alpha}$ and leptons $l_i$ which are responsible
for the decays (\ref{eq:KL0B0Bsdecays}) and induced by the gauge
group (\ref{eq:Gc4L4R})    
can be written in general case as~\cite{Smirnov:2007hv, Smirnov:2008zzb}
\begin{eqnarray}
  \emph{L}_{V^{LR}dl} &=& \frac{g^L_4}{\sqrt{2}} (\bar{d}_{p \alpha}
[ (K^L_2)_{pi}\gamma^{\mu}P_L] l_i)V^L_{\alpha \mu} +
\frac{g^R_4}{\sqrt{2}} (\bar{d}_{p \alpha}
[ (K^R_2)_{pi}\gamma^{\mu}P_R] l_i)V^R_{\alpha \mu} + h.c. , \hspace{5mm}
\label{eq:lagrVLRdl}
\end{eqnarray}
where $g^{L}_4$, $g^{R}_4$ are the gauge coupling constants
of the group (\ref{eq:Gc4L4R}),   
$p, i = 1,2,3 $ are the quark and lepton generation idexes,
$ \alpha = 1, 2, 3 $ is the $ SU_{c}(3) $ colour index,
$d_p=(d,s,b)$, $l_{i}=(e, \mu, \tau)$ are down quarks and leptons,
$P_{L,R}=(1\pm \gamma_5)/2$ are the left and right operators of fermions
and $K^{L,R}_2$ are the unitary matricies which describe the mixing
of down fermion in the leptoquark currents. In general case the interaction
of the leptoquarks with fermions contains four mixing matrices $K^{L,R}_a$
for up$(a=1)$  and down$(a=2)$ quarks and leptons.
These matrices are specific for the models with the four color quark-lepton
symmetry.
Some details concerning the mixing matrices $K^{L,R}_a$ can be found in
Refs.~\cite{Smirnov:1995jq,AD22,Smirnov:2007hv,Smirnov:2008zzb,Smirnov:2018ske}.

The group  (\ref{eq:Gc4L4R})    
after breaking down to the $ SU_{c}(3) $
colour group reproduces in particular the usual QCD interaction and as a result
the gauge coupling constants $g^{L}_4$, $g^{R}_4$ should satisfy the relation
\begin{eqnarray}
g^L_4 g^R_4 / \sqrt{(g^L_4)^2+(g^R_4)^2} = g_{st}(M_c) ,
\label{eq:gLgRgst}
\end{eqnarray}
where $g_{st}(M_c)$ is the strong coupling constant at the mass scale $M_c$
of the chiral four color symmetry (\ref{eq:Gc4L4R}).   

The branching ratios of the decays of pseudoscalar mesons
$P =( K^0_L, \, B^0, \, B_s )$
into lepton-antilepton pairs of type (\ref{eq:KL0B0Bsdecays})
induced by the chiral gauge leptoquarks $V^L, \, V^R$
can be presented in the form
\begin{eqnarray}
&& \hspace{-15mm} Br_{V^{LR}}(P \to l^+_i l^-_i) =
B_{Pl} \, \sqrt{1-4 m_{l_i}^2/m_P^2  } \,\, \beta_{P, \, i i}^2, \hspace{5mm}
\mbox{for} \hspace{3mm} l = l_i = e, \, \mu, \, \tau,
\label{eq:Br_VLRPlplm}
\\
&& \hspace{-15mm} Br_{V^{LR}}(P \to  l \, l^{\prime}) =
B_{Pl} \,( 1 - m_{l}^2/m_P^2 )^2   \,\, \beta_{P, \, ll^{\prime}}^2, \hspace{5mm}
\mbox{for} \hspace{3mm} l \, l^{\prime} = \mu e, \, \tau e, \, \tau \mu,
\label{eq:Br_VLRPll1}
\end{eqnarray}
where $B_{Pl}$ are the typical branching ratios of these decays
and $\beta_{P, \, i i}^2$, $\beta_{P, \, ll^{\prime}}^2$ are the mixing factors
depending on the mixing matrices $K^{L}_2$, $K^{R}_2$
and on the masses and coupling constants of leptoquarks $V^L, V^R$.

The branching ratio $B_{Pl}$ in eqs. (\ref{eq:Br_VLRPlplm}), (\ref{eq:Br_VLRPll1})
can be written as
\begin{eqnarray}
&&  B_{Pl} =
\frac{m_{P} \, \pi \,\alpha^2_{st}(M_c) \, f_{P}^2 \, m_l^2 \, }
{2 \,(\bbar{m}_{V})^4 \, \Gamma_P^{tot}  } ,
\label{eq:BPl}
\end{eqnarray}
where $m_{P}$, $m_l$  are the masses of $P$ meson and lepton,
$f_{P}$ is the form factor parametrizing the matrix elements
of the axial and pseudoscalar quark currents of $P$ meson in the standard way,
$\bbar{m}_{V}$ is some typical leptoquark mass (conveniently the mass of
the lightest leptoquark $m_{V^L}$ or $m_{V^R}$)
and $\Gamma_P^{tot}$ is the total width of $P$ meson.

The branching ratios $Br_{V^{LR}}(P \to  l \, l^{\prime})$ in eq.(\ref{eq:Br_VLRPll1})
denote the sums of the branching ratios of the charge conjugated final states
\begin{eqnarray}
&&
\hspace{-15mm}Br_{V^{LR}}(P \to \mu e) =
Br_{V^{LR}}(P \to \mu^+ e^-) + Br_{V^{LR}}(P \to e^+ \mu^-),
\label{eq:Br_VLRPmue}
\\
&&
\hspace{-15mm}Br_{V^{LR}}(P \to \tau e) =
Br_{V^{LR}}(P \to \tau^+ e^-) + Br_{V^{LR}}(P \to e^+ \tau^-),
\\
\label{eq:Br_VLRPtaue}
&&
\hspace{-15mm}Br_{V^{LR}}(P \to \tau \mu) =
Br_{V^{LR}}(P \to \tau^+ \mu^-) + Br_{V^{LR}}(P \to \mu^+ \tau^-)
\label{eq:Br_VLRPtaumu}
\end{eqnarray}
with $m_l$ being the mass of heavyest lepton ($m_l > m_{l^{\prime}}$)
and with account of the relations
\begin{eqnarray}
&& \hspace{-15mm}
 m_{\tau} \gg m_{\mu} \gg m_e.
\label{eq:mtauggmmuggme}
\end{eqnarray}
According to the definitions (\ref{eq:Br_VLRPmue})--(\ref{eq:Br_VLRPtaumu})
the mixing factors $\beta_{P, \, ll^{\prime}}^2$ in eq.(\ref{eq:Br_VLRPll1})
can be written as
\begin{eqnarray}
&&
\beta_{P,\, \mu e}^2 = \beta_{P,\, 21}^2 + \beta_{P,\, 12}^2  , \hspace{8mm}
\hspace{-4mm}  \beta_{P,\, \tau e}^2 =
\beta_{P,\, 31}^2 +\beta_{P,\, 13}^2 , \hspace{3mm}
\beta_{P,\, \tau \mu}^2 = \beta_{P,\, 32}^2 +\beta_{P,\, 23}^2 . \hspace{3mm}
\label{eq:betaVLR2Pll1b}
\end{eqnarray}
With branching ratios (\ref{eq:Br_VLRPlplm})--(\ref{eq:BPl})
the mixing factors $\beta_{P,i j}^2$
in eqs.(\ref{eq:Br_VLRPlplm}), (\ref{eq:betaVLR2Pll1b})
with account of the quark content of $P$ meson
depend on the matrix elements
$(K^{L,R}_2)_{pi}, \, (K^{L,R}_2)_{qj}$
of the mixing matrices $K^{L,R}_2$ and on the relations between the masses
and coupling constants of the leptoquarks $V^L, V^R$.

Each of the unitary $3\times3$ mixing matrices $K^{L,R}_2$
in general case can be parametrized by three angles
$\theta^{L,R}_{12}, \, \theta^{L,R}_{23}, \, \theta^{L,R}_{13}$
and six phases
$\delta^{L,R}, \, \varepsilon^{L,R}, \, \varphi^{L,R}_0, \, \varphi^{L,R}_1, \,
\varphi^{L,R}_2, \, \varphi^{L,R}_3 $
as \cite{Smirnov:2018ske}
\begin{eqnarray}
K^{L,R}_2 \hspace{-1mm} = e^{i\varphi^{L,R}_0} \hspace{-1mm}  \left( \begin{array}{ccc}
 c_{12}c_{13}e^{i\varphi_1} & s_{12}c_{13}e^{i\varphi_2} & s_{13}e^{i\varphi_3}   \\
\hspace{-3mm}   (-s_{12}c_{23}-c_{12}s_{23}s_{13}e^{i\delta})e^{i\varphi_{21}}   &
\hspace{-3mm}  (c_{12}c_{23}-s_{12}s_{23}s_{13}e^{i\delta})e^{i\varphi_{22}} &
\hspace{-3mm}  s_{23}c_{13}e^{i\varphi_{23}}  \\
\hspace{-3mm}   (s_{12}s_{23}-c_{12}c_{23}s_{13}e^{i\delta})e^{i\varphi_{31}} &
\hspace{-3mm}  (-c_{12}s_{23}-s_{12}c_{23}s_{13}e^{i\delta})e^{i\varphi_{32}}  &
\hspace{-3mm}  c_{23}c_{13}e^{i\varphi_{33}} 
\end{array} \hspace{-2mm}  \right)^{\hspace{-2mm} L,R},
\label{eq:KLR2}
\end{eqnarray}
where
\begin{eqnarray}
&&
\hspace{-15mm} \varphi^{L,R}_{21} = ( \varphi_1+\varepsilon)^{L,R} , \hspace{5mm}
\varphi^{L,R}_{22} =  (\varphi_2+\varepsilon)^{L,R},  \hspace{5mm}
\varphi^{L,R}_{23} =  (\varphi_3+\delta+\varepsilon)^{L,R},
\nonumber
\label{eq:phi2i}
\\
&&
\hspace{-15mm} \varphi^{L,R}_{31} =
 -(  \varphi_2 \hspace{-1mm} + \hspace{-1mm} \varphi_3 \hspace{-1mm} +
 \hspace{-1mm} \delta \hspace{-1mm} + \hspace{-1mm} \varepsilon )^{L,R},  \hspace{2mm}
\varphi^{L,R}_{32} = -(  \varphi_1 \hspace{-1mm} + \hspace{-1mm} \varphi_3 \hspace{-1mm}
 + \hspace{-1mm} \delta \hspace{-1mm} + \hspace{-1mm} \varepsilon )^{L,R},  \hspace{2mm}
\varphi^{L,R}_{33} = -(  \varphi_1 \hspace{-1mm} + \hspace{-1mm} \varphi_2 \hspace{-1mm}
 + \hspace{-1mm} \varepsilon )^{L,R},
\nonumber
\label{eq:phi3i}
\\
&&
\hspace{-15mm}
s^{L,R}_{ij} = \sin \theta^{L,R}_{ij}, \, c^{L,R}_{ij} = \cos \theta^{L,R}_{ij}.
\nonumber
\end{eqnarray}

The use of the mixing matrices $K^{L,R}_2$ in the form (\ref{eq:KLR2})
gives for the mixing factors $\beta_{P,i j}^2$ the next expressions
for $P=K^0_L$
\begin{eqnarray}
&&
\beta_{K^0_L, \, 11}^2 = \frac{1}{8}
\bigg [ \,
\frac{\mu_L^2 }{\cos^2{\gamma_L}} \, {c}_{12}^{L} {c}_{13}^{L}
\Big( e^{i \varepsilon ^L}
\big (
{s}_{12}^{L} {c}_{23}^{L} + e^{i \delta ^L} {c}_{12}^{L} {s}_{13}^{L} {s}_{23}^{L}
\big )
+ c.c.
\Big) +
L \leftrightarrow R \,
\bigg ] ^2 ,
\label{eq:betaVLR2K0L11}
\\
&&
\beta_{K^0_L, \, 22}^2 = \frac{1}{8}
\bigg [ \,
\frac{\mu_L^2 }{\cos^2{\gamma_L}} \, {s}_{12}^{L} {c}_{13}^{L}
\Big (
e^{i \varepsilon ^L}
\big (
{c}_{12}^{L} {c}_{23}^{L} -
e^{i \delta ^L} {s}_{12}^{L}{s}_{13}^{L} {s}_{23}^{L}
\big )
+ c.c.
\Big ) +
L \leftrightarrow R \,
\bigg ] ^2 ,
\\
\label{eq:betaVLR2K0L22}
&&
\beta_{K^0_L, \, 21}^2 = \beta_{K^0_L, \, 12}^2 = \frac{1}{16}
\bigg [ \,
\frac{\mu_L^4 }{\cos^4{\gamma_L}}
\big ( \, c_{13}^L \big )^2 \,
\Big\vert \,
{c}_{12}^{L} e^{i {\varepsilon }^L}
\big (
{c}_{12}^{L} {c}_{23}^{L} - e^{i {\delta }^L} {s}_{12}^{L} {s}_{13}^{L} {s}_{23}^{L}
\big )
-
\nonumber
\\
&&
\hspace{17mm}
- {s}_{12}^{L} e^{-i {\varepsilon }^L}
\big (
{c}_{23}^{L} {s}_{12}^{L} + e^{-i {\delta }^L} {c}_{12}^{L} {s}_{13}^{L} {s}_{23}^{L}
\big )
\, \Big\vert ^2 + L \leftrightarrow R \,
\bigg ] ,
\label{eq:betaVLR2K0L2112}
\end{eqnarray}
for $P=B^0$
\begin{eqnarray}
&&
\beta_{B^0, \, 11}^2 = \frac{1}{4}
\Big\vert \,
\frac{\mu_L^2 }{\cos^2{\gamma_L}} \,
{c}_{12}^{L} {c}_{13}^{L} \, e^{i \chi ^L + i \varepsilon ^L}
\big (
{s}_{12}^{L} {s}_{23}^{L}e^{i \delta ^L} - {c}_{12}^{L} {s}_{13}^{L} {c}_{23}^{L}
\big )
 + L \leftrightarrow R \,
\Big\vert ^2 ,
\label{eq:betaVLR2B011}
\\
&&
\beta_{B^0, \, 22}^2 = \frac{1}{4}
\Big\vert \,
\frac{\mu_L^2 }{\cos^2{\gamma_L}} \,
{s}_{12}^{L}{c}_{13}^{L} \, e^{i \chi ^L + i \varepsilon ^L}
\big (
{c}_{12}^{L} {s}_{23}^{L}e^{i \delta ^L} + {s}_{12}^{L} {s}_{13}^{L} {c}_{23}^{L}
\big )
+L \leftrightarrow R \,
\Big\vert ^2 ,
\label{eq:betaVLR2B022}
\\
&&
\beta_{B^0, \, 33}^2 = \frac{1}{4}
\Big\vert  \,
\frac{\mu_L^2 }{\cos^2{\gamma_L}}  \,
{c}_{13}^{L} {s}_{13}^{L} {c}_{23}^{L}  \, e^{i \chi ^L + i \varepsilon ^L}
+ L \leftrightarrow R \,
\Big\vert ^2 ,
\label{eq:betaVLR2B033}
\\
&&
\beta_{B^0, \, 21}^2 =\frac{1}{8}
\bigg [ \,
\frac{\mu_L^4 }{\cos^4{\gamma_L}} \,
\big ( c_{12}^L \big )^2 \big ( c_{13}^L \big )^2 \,
\Big\vert  \,
{s}_{12}^{L} {s}_{13}^{L} {c}_{23}^{L} + {c}_{12}^{L} {s}_{23}^{L} e^{i \delta ^L}
\Big\vert ^2
+ L \leftrightarrow R \,
\bigg ] ,
\label{eq:betaVLR2B021}
\\
&&
\beta_{B^0, \, 12}^2 = \frac{1}{8}
\bigg [ \,
\frac{\mu_L^4 }{\cos^4{\gamma_L}} \,
\big ( c_{13}^L \big )^2 \big ( s_{12}^L \big )^2 \,
\Big\vert  \,
{s}_{12}^{L} {s}_{23}^{L} - {c}_{12}^{L} {s}_{13}^{L} {c}_{23}^{L} e^{i \delta ^L}
\Big\vert
^2 + L \leftrightarrow R \,
\bigg ] ,
\label{eq:betaVLR2B012}
\\
&&
\beta_{B^0, \, 31}^2 = \frac{1}{8}
\bigg [ \,
\frac{\mu_L^4 }{\cos^4{\gamma_L}} \,
\big ( c_{12}^L\big )^2 \big ( c_{13}^L\big )^4 \big ( c_{23}^L\big )^2
+L \leftrightarrow R \,
\bigg ] ,
\label{eq:betaVLR2B031}
\\
&&
\beta_{B^0, \, 13}^2 = \frac{1}{8}
\bigg [ \,
\frac{\mu_L^4 }{\cos^4{\gamma_L}} \,
\big ( s_{13}^L \big )^2 \,
\Big\vert \, {s}_{12}^{L} {s}_{23}^{L} e^{i \delta ^L}-
{c}_{12}^{L} {s}_{13}^{L} {c}_{23}^{L}
\Big\vert ^2+L \leftrightarrow R \,
\bigg ] ,
\label{eq:betaVLR2B013}
\\
&&
\beta_{B^0, \, 32}^2 =\frac{1}{8}
\bigg [ \,
\frac{\mu_L^4 }{\cos^4{\gamma_L}} \,
\big ( c_{13}^L \big ) ^4 \big ( c_{23}^L \big ) ^2 \big ( s_{12}^L \big ) ^2 +
L \leftrightarrow R \,
\bigg ] ,
\\
\label{eq:betaVLR2B032}
&&
\beta_{B^0, \, 23}^2 = \frac{1}{8}
\bigg [ \,
\frac{\mu_L^4 }{\cos^4{\gamma_L}} \,
\big ( s_{13}^L \big ) ^2 \, \Big\vert \, {s}_{12}^{L} {s}_{13}^{L} {c}_{23}^{L} +
{c}_{12}^{L} {s}_{23}^{L} e^{i \delta ^L} \Big\vert ^2 + L \leftrightarrow R \,
\bigg ] ,
\label{eq:betaVLR2B023}
\end{eqnarray}
and for $P=B_s$
\begin{eqnarray}
&&
\hspace{-15mm}
\beta_{B_s, \, 11}^2 = \frac{1}{4}
\Big\vert \, \frac{\mu_L^2 }{\cos^2{\gamma_L}} \,
e^{i \chi ^L + 2 i \varepsilon ^L}
 \big (\hspace{-1mm}-\hspace{-1mm}{c}_{12}^{L} {s}_{13}^{L} {c}_{23}^{L}
\hspace{-1mm}+\hspace{-1mm}{s}_{12}^{L} {s}_{23}^{L}
e^{i \delta ^L} \big )  \big ( {s}_{12}^{L} {c}_{23}^{L}\hspace{-1mm}+\hspace{-1mm}
{c}_{12}^{L} {s}_{13}^{L} {s}_{23}^{L} e^{i \delta ^L} \big ) +
L \leftrightarrow R \, \Big\vert ^2 ,
\label{eq:betaVLR2Bs11}
\\
&&
\hspace{-15mm}
\beta_{B_s, \, 22}^2 = \frac{1}{4}
\Big\vert \, \frac{\mu_L^2 }{\cos^2{\gamma_L}}  \, e^{i \chi ^L + 2 i \varepsilon ^L}
 \big ( {s}_{12}^{L} {s}_{13}^{L} {c}_{23}^{L}+{c}_{12}^{L} {s}_{23}^{L}
e^{i \delta ^L} \big )
 \big ( {c}_{12}^{L} {c}_{23}^{L}-{s}_{12}^{L} {s}_{13}^{L} {s}_{23}^{L} e^{i \delta ^L}
 \big ) +
L \leftrightarrow R \,\Big\vert  ^2 ,
\label{eq:betaVLR2Bs22}
\\
&&
\hspace{-15mm}
\beta_{B_s, \, 33}^2 = \frac{1}{4}
\Big\vert \, \frac{\mu_L^2 }{\cos^2{\gamma_L}} \,
 \big ( c_{13}^L \big ) ^2 {c}_{23}^{L} {s}_{23}^{L}
e^{i \chi ^L +i \delta ^L + 2 i \varepsilon ^L}
 + L \leftrightarrow R \,\Big\vert ^2 ,
\label{eq:betaVLR2Bs33}
\\
&&
\hspace{-15mm}
\beta_{B_s, \, 21}^2 =\frac{1}{8}
\bigg [ \,
\frac{\mu_L^4 }{\cos^4{\gamma_L}}\Big\vert {c}_{12}^{L} {s}_{23}^{L}+
{s}_{12}^{L} {s}_{13}^{L} {c}_{23}^{L} e^{i \delta ^L}\Big\vert ^2
\Big\vert {s}_{12}^{L} {c}_{23}^{L}+{c}_{12}^{L} {s}_{13}^{L} {s}_{23}^{L}
e^{i \delta ^L}\Big\vert ^2+L \leftrightarrow R \,\bigg ] ,
\label{eq:betaVLR2Bs21}
\\
&&
\hspace{-15mm}
\beta_{B_s, \, 12}^2 = \frac{1}{8}
\bigg [ \,
\frac{\mu_L^4 }{\cos^4{\gamma_L}}\Big\vert {s}_{12}^{L} {s}_{23}^{L}-
{c}_{12}^{L} {s}_{13}^{L} {c}_{23}^{L} e^{i \delta ^L}\Big\vert ^2
\Big\vert {c}_{12}^{L} {c}_{23}^{L}-{s}_{12}^{L} {s}_{13}^{L} {s}_{23}^{L}
e^{i \delta ^L}\Big\vert ^2+L \leftrightarrow R \,
\bigg ] ,
\label{eq:betaVLR2Bs12}
\\
&&
\hspace{-15mm}
\beta_{B_s, \, 31}^2 = \frac{1}{8}
\bigg [ \,
\frac{\mu_L^4 }{\cos^4{\gamma_L}}
 \big ( c_{13}^L \big )^2 \big ( c_{23}^L \big )^2
\Big\vert {s}_{12}^{L} {c}_{23}^{L}+{c}_{12}^{L} {s}_{13}^{L} {s}_{23}^{L}
e^{i \delta ^L}\Big\vert ^2+L \leftrightarrow R \,
\bigg ] ,
\label{eq:betaVLR2Bs31}
\\
&&
\hspace{-15mm}
\beta_{B_s, \, 13}^2 = \frac{1}{8}
\bigg [ \,
\frac{\mu_L^4 }{\cos^4{\gamma_L}}
 \big ( c_{13}^L\big )^2 \big ( s_{23}^L \big )^2\Big\vert {s}_{12}^{L} {s}_{23}^{L}
e^{i \delta ^L}-{c}_{12}^{L} {s}_{13}^{L} {c}_{23}^{L}\Big\vert ^2
+ L \leftrightarrow R \,
\bigg ] ,
\label{eq:betaVLR2Bs13}
\\
&&
\hspace{-15mm}
\beta_{B_s, \, 32}^2 = \frac{1}{8}
\bigg [ \,
\frac{\mu_L^4 }{\cos^4{\gamma_L}}
 \big ( c_{13}^L\big )^2 \big ( c_{23}^L\big )^2
\Big\vert {c}_{12}^{L} {c}_{23}^{L}-{s}_{12}^{L} {s}_{13}^{L} {s}_{23}^{L}
e^{i \delta ^L}\Big\vert ^2 + L \leftrightarrow R \,
\bigg ] ,
\label{eq:betaVLR2Bs32}
\\
&&
\hspace{-15mm}
\beta_{B_s, \, 23}^2 = \frac{1}{8}
\bigg [ \,
\frac{\mu_L^4 }{\cos^4{\gamma_L}}
\big ( c_{13}^L\big )^2\big ( s_{23}^L\big )^2
\Big\vert {c}_{12}^{L} {s}_{23}^{L}+{s}_{12}^{L} {s}_{13}^{L} {c}_{23}^{L}
e^{i \delta ^L} \Big\vert ^2  + L \leftrightarrow R \,
\bigg ] ,
\label{eq:betaVLR2Bs23}
\end{eqnarray}
where \,
$\mu_{L,\,R} = \bbar{m}_{V}/m_{V^{L,\,R}} $, \,
$\gamma_{L,R}$ are defined
through
$\sin^2{\gamma_{L,R}} = (g^{L,R}_4)^2/\big [ (g^{L}_4)^2 + (g^{R}_4)^2 \big ]$
so that
%
\begin{eqnarray}
&&
\hspace{-15mm}
%
\cos^2{\gamma_{L,R}} = g_{st}^2(M_c)/(g^{L,R}_4)^2
 = \alpha_{st}(M_c)/\alpha^{L,R}_{4} \, , \hspace{5mm} 
\cos^2{\gamma_{L}}  + \cos^2{\gamma_{R}} =1
\label{eq:cosgammaLR}
\end{eqnarray}
with $g_{st}(M_c)$ beeng defined by the relation (\ref{eq:gLgRgst})
and
$\chi ^{L,R} = {\varphi }_1^{L,R} + {\varphi }_2^{L,R} + {\varphi }_3^{L,R} $.

The mixing factors~$\beta_{P, \, i i}^2$, (\ref{eq:betaVLR2Pll1b}),
(\ref{eq:betaVLR2K0L11})--(\ref{eq:betaVLR2Bs23})
describe in the general form
the effect of the fermion mixing in leptoquark currents
in the case (\ref{eq:mtauggmmuggme}) on the branching
ratios~(\ref{eq:Br_VLRPlplm}),~(\ref{eq:Br_VLRPll1})
of the decays of pseudoscalar
mesons $P =( K^0_L, \, B^0, \, B_s )$ into lepton-antilepton pairs
induced by the chiral gauge leptoquarks.
These mixing factors in general case depend on six angles
$\theta^{L,R}_{12}, \, \theta^{L,R}_{23}, \, \theta^{L,R}_{13}$
and six phases
$\delta^{L,R}, \, \varepsilon^{L,R}, \, \chi ^{L,R} $
and can be used for the analysis of the branching
ratios 
of these decays in dependence on the mixing
angles and phases of the mixing matrices (\ref{eq:KLR2}).
With account of the expressions (\ref{eq:betaVLR2B011})--(\ref{eq:betaVLR2Bs23})
one can see that for $P= ( B^0, \,B_s ) $ the mixing factors~$\beta_{P, \, i i}^2$
and (\ref{eq:betaVLR2Pll1b}) in the case of $\mu_{R\,(L)} = 0 $ satisfy the relations
\begin{eqnarray}
&& \hspace{-15mm}
\sum_{i=1}^3 \beta_{P, \, i i}^2 + 2 \, \Big (
\beta_{P, \, \mu e}^2 + \beta_{P, \, \tau e}^2 + \beta_{P, \, \tau \mu}^2 \Big ) =
\frac{1}{4} \, \frac{\mu_{L(R)}^4 }{\cos^4{\gamma_{L(R)}}} \, ,
\label{eq:sumbeta2}
\nonumber
\end{eqnarray}
which is a result of the relations (\ref{eq:mtauggmmuggme}) and
of unitarity of the matrices (\ref{eq:KLR2}).

Further we will consider the special case that one of two leptoquarks is
much more heavy that the another so that its contribution to decays under consideration
can be neglected. For definiteness we assume that
\begin{eqnarray}
&& \hspace{-15mm}
m_{V^R} \gg m_{V^L}
\label{eq:mVRggmVL}
\end{eqnarray}
and for account of the contribution only of the lightest leptoquark $V^L$
it is sufficient in the expressions (\ref{eq:BPl}),
(\ref{eq:betaVLR2K0L11})--(\ref{eq:betaVLR2Bs23})
to set
$\bbar{m}_{V} =m_{V^L}, \, \mu_L=1, \, \mu_R = 0. $
In this case the mixing factors
(\ref{eq:betaVLR2K0L11})--(\ref{eq:betaVLR2K0L2112})
for $K^0_L$ meson still depend on three angles
$\theta^{L}_{12}, \, \theta^{L}_{23}, \, \theta^{L}_{13}$ and on two
phases $\delta^{L}, \, \varepsilon^{L} $
whereas the mixing factors
(\ref{eq:betaVLR2B011})--(\ref{eq:betaVLR2Bs23})
for $B^0 , \, B_s$ mesons depend on these three angles
and on the only phase $\delta^{L}$.
The opposite case of $m_{V^L} \gg m_{V^R}$ can be obtained from
the formulas given below by the simple exchange $L \leftrightarrow R$.

We have numerically analysed the branching
ratios (\ref{eq:Br_VLRPlplm})--(\ref{eq:BPl}) with mixing factors
$ \beta_{P, \, i i}^2,$ and $ \beta_{P, \, ll^{\prime}}^2 $
defined by the
equations (\ref{eq:betaVLR2Pll1b}), (\ref{eq:betaVLR2K0L11})--(\ref{eq:betaVLR2Bs23})
with account of experimental data on the decays~(\ref{eq:KL0B0Bsdecays}).
%
%
The experimental data on the branching ratios $Br(P \to l^+_i l^-_i)^{exp}$,
$Br(P \to l \, l^{\prime})^{exp}$ are taken from the Ref. \cite{Tanabashi:2018oca2019}
except the branching ratios of the decays $ B^0, B_s  \to \tau\mu$
for which we use the recent data \cite{Aaij:2019okb}
\begin{eqnarray}
&&  
Br(B^0 \to \tau \mu)^{exp} < 1.4\cdot 10^{-5}  , 
\label{eq:BrB0taumuexp}
\\
&&  
Br(B_s \to \tau \mu)^{exp} < 4.2\cdot 10^{-5} .   
\label{eq:BrBstaumuexp}
\end{eqnarray}
%
%
The experimental data 
under consideration are presented
in the second column of the Table~\ref{tab:BrVLPl1l2}.
In the third column of this table 
we present for comparison the SM predictions
$Br_{SM}(P\to l_1 l_2)$
for the branching ratios of the diagonal leptonic decays
$K^0_L \to e^+ e^-$~\cite{Valencia_NP1998, GDP_PRL1998},
$K^0_L \to \mu^+ \mu^-$~\cite{Tanabashi:2018oca} and
$B_s, B^0 \to l^+_i l^-_i$~\cite{Bobeth:2013uxa}. 
We vary the mixing angles $\theta^{L}_{12}, \, \theta^{L}_{23}, \, \theta^{L}_{13}$ and
phases $\delta^{L}, \, \varepsilon^{L} $
to find the minimal chiral gauge leptoquark mass $m_{V^L}$ satisfying
these experimental data with account that the SM contributions to some of the diagonal
leptonic decays are dominant.

As seen from the Table~\ref{tab:BrVLPl1l2} the  experimental branching ratios of
the diagonal $K^0_L \to e^+ e^-, \mu^+ \mu^-$ decays are almost completely saturated
by the (long-distance) SM contributions so that the new physics contributions
to these decays should be small. With account also  of the very stringent experimental
limit on the branching ratio of the nondiagonal $K^0_L \to \mu e $ decay
the possible leptoquark contributions to the branching ratios
of $K^0_L \to e^+ e^-, \mu^+ \mu^-, \mu e  $ decays should be small.
It means that for the small leptoquark mass
the mixing factors $\beta_{K^0_L, \, i j }^2$ must be very small (close to zero).
Below we believe for definitness that
\begin{eqnarray}
&&
\hspace{-15mm}
\beta_{K^0_L, \, i j }^2 = 0.
\label{eq:betaK0Lijto0}
\end{eqnarray}

The measured values $Br(B_s, B^0 \to \mu^+ \mu^-)^{exp}$ are consistent
with the corresponding SM predictions
within the experimental and theoretical uncertainties
with the sufficiently good degree of this consistency in the case
of $B_s \to \mu^+ \mu^-$ decay.
We assume below that the experimental branching ratios
$Br(B_s, B^0 \to \mu^+ \mu^-)^{exp}$ of these decays are saturated mainly
by the SM contributions so that the contributions
$ Br_{V^L}(B_s, B^0 \to \mu^+ \mu^-)$ of the chiral gauge
leptoquark~$V^L$ to these branching ratios can be assumed to be small.
In this case the mixing factors $\beta_{B_s, \, 2 2 }^2$,   $\beta_{B^0, \, 2 2 }^2$
for the small leptoquark mass should be sufficiently small.
Below we assume for definitness that
\begin{eqnarray}
&&
\hspace{-15mm}
\beta_{B_s, \, 2 2 }^2 = 0 ,  \hspace{5mm}       \beta_{B^0,  \, 2 2 }^2 = 0,
\label{eq:betaBsB022to0}
\end{eqnarray}
that is $ Br_{V^L}(B_s, B^0 \to \mu^+ \mu^-) = 0$,
and find the minimal chiral gauge leptoquark mass~$m_{V^L}$ satisfying
the experimental data with account of
the conditions (\ref{eq:betaK0Lijto0}), (\ref{eq:betaBsB022to0}).

We have found three solutions of the equations (\ref{eq:betaK0Lijto0}):
\begin{eqnarray}
&&\hspace{-15mm}
\mbox{solution 1:} \hspace{10mm}
\theta^{L}_{23}= \pi /2, \hspace{5mm}
\delta^{L} + \varepsilon^{L}  = 3 \pi/2 ,
\label{eq:solution0}
\\
&&\hspace{-15mm}
\mbox{solution 2:} \hspace{10mm}
\theta^{L}_{13}= \pi /2,
\label{eq:solution02}
\\
&&\hspace{-15mm}
\mbox{solution 3:} \hspace{10mm}
\theta^{L}_{23}= \pi /2, \hspace{5mm}
\theta^{L}_{13}= 0 .
\label{eq:solution03}
\end{eqnarray}
The solution (\ref{eq:solution0}) contains two free independent angles
$\theta^{L}_{12}, \, \theta^{L}_{13}$ whereas each of the
other ones
contains only one angle (in the case (\ref{eq:solution02}) two angles
$\theta^{L}_{23}, \, \theta^{L}_{13}$ form one effective angle~$\bar{\theta}^{L}$
defined as
$\sin \bar{\theta}^{L} =  | c_{23}^{L}c_{12}^{L} - s_{23}^{L}s_{12}^{L}e^{i\delta^{L}} |$ ).
The solution (\ref{eq:solution0}) results in the more lower mass limits for
the chiral gauge leptoquarks in comparision with the solutions
(\ref{eq:solution02}), (\ref{eq:solution03})
and we use it
in the further analysis.

In the case (\ref{eq:mVRggmVL}), (\ref{eq:solution0}) the mixing factors
$ \beta_{P, \, i i}^2,$ and $ \beta_{P, \, ll^{\prime}}^2 $
defined by the
equations (\ref{eq:betaVLR2Pll1b}), (\ref{eq:betaVLR2K0L11})--(\ref{eq:betaVLR2Bs23})
are simplifyed and the nonzero mixing factors take the form
for $P=B^0$
\begin{eqnarray}
&&
\beta_{B^0, \, 11}^2 = \beta_{B^0, \, 22}^2 = \frac{1}{4}
\frac{ 1 }{\cos^4{\gamma_L}} \,
\big ( {c}_{12}^{L} {s}_{12}^{L} {c}_{13}^{L} \big )^2
,
\label{eq:betaVLR2B011s0}
\\
&&
\beta_{B^0, \, \mu e}^2 =\frac{1}{8}
\frac{ 1 }{\cos^4{\gamma_L}} \,
\big [ \, \big ( c_{12}^L \big )^4 + \big ( s_{12}^L \big )^4 \, \big ]
\big ( c_{13}^L \big )^2 ,
\label{eq:betaVLR2B0mues0}
\\
&&
\beta_{B^0, \, \tau e }^2 = \frac{1}{8}
\frac{ 1 }{\cos^4{\gamma_L}} \,
\big ( s_{12}^L \big )^2 \big ( {s}_{13}^{L}\big )^2  ,
\label{eq:betaVLR2B0taues0}
\\
&&
\beta_{B^0, \, \tau \mu}^2 = \frac{1}{8}
\frac{ 1 }{\cos^4{\gamma_L}} \,
\big ( c_{12}^L \big ) ^2 \big ( s_{13}^L \big ) ^2 ,
\label{eq:betaVLR2B0taumus0}
\end{eqnarray}
and for $P=B_s$
\begin{eqnarray}
&&
\beta_{B_s, \, 11}^2 = \beta_{B_s, \, 22}^2 = \frac{1}{4}
\frac{ 1 }{\cos^4{\gamma_L}} \,
\big ( {c}_{12}^{L} {s}_{12}^{L} {s}_{13}^{L} \big )^2,
\label{eq:betaVLR2Bs11s0}
\\
&&
\beta_{B_s, \, \mu e}^2 =\frac{1}{8}
\frac{ 1 }{\cos^4{\gamma_L}} \,
 \big [ \, \big ( c_{12}^L \big ) ^4 +  \big ( s_{12}^L \big ) ^4 \, \big ]
\big ( s_{13}^L \big ) ^2 ,
\label{eq:betaVLR2Bsmues0}
\\
&&
\beta_{B_s, \, \tau e}^2 = \frac{1}{8}
\frac{ 1 }{\cos^4{\gamma_L}} \,
\big ( s_{12}^L \big ) ^2 \big ( c_{13}^L \big ) ^2 ,
\label{eq:betaVLR2Bstaues0}
\\
&&
\beta_{B_s, \, \tau \mu}^2 = \frac{1}{8}
\frac{ 1 }{\cos^4{\gamma_L}} \,
\big ( c_{12}^L \big ) ^2 \big ( c_{13}^L \big ) ^2  .
\label{eq:betaVLR2Bstaumus0}
\end{eqnarray}

With accound of expessions (\ref{eq:betaVLR2B011s0}), (\ref{eq:betaVLR2Bs11s0})
the equations (\ref{eq:betaBsB022to0}) give   %
\begin{eqnarray}
&&\hspace{-15mm}
\theta^{L}_{12} = 0 \,\, (\pi/2) 
\label{eq:theta120pi2}
\end{eqnarray}
and the nonzero mixing factors in (\ref{eq:betaVLR2B011s0})-(\ref{eq:betaVLR2Bstaumus0})
take the form
\begin{eqnarray}
&&
\beta_{B^0, \, \mu e}^2 =\frac{1}{8}
\frac{ 1 }{\cos^4{\gamma_L}} \,
\big ( c_{13}^L \big )^2 ,
\label{eq:betaVLR2B0mues0thete120pi2}
\\
&&
\beta_{B_s, \, \mu e}^2 =\frac{1}{8}
\frac{ 1 }{\cos^4{\gamma_L}} \,
\big ( s_{13}^L \big ) ^2 ,
\label{eq:betaVLR2Bsmues0thete120pi2}
\\
&&
\beta_{B^0, \, \tau \mu}^2,  \,\, (\beta_{B^0, \, \tau e }^2 ) = \frac{1}{8}
\frac{ 1 }{\cos^4{\gamma_L}} \,
\big ( s_{13}^L \big ) ^2 \hspace{5mm} \mbox{for} \,\, \theta^{L}_{12} = 0, \,\, (\pi/2),  
\label{eq:betaVLR2B0taumutaues0thete120pi2}
\\
&&
\beta_{B_s, \, \tau \mu}^2,  \,\,(\beta_{B_s, \, \tau e}^2 ) = \frac{1}{8}
\frac{ 1 }{\cos^4{\gamma_L}} \,
\big ( c_{13}^L \big ) ^2 \hspace{5mm} \mbox{for} \,\, \theta^{L}_{12} = 0, \,\, (\pi/2).
\label{eq:betaVLR2Bstaumutaues0thete120pi2}
\end{eqnarray}

In the numerical analysis we use the data on the masses
of leptons and quarks and the data on the masses
and life times $\tau_{P}$ ($\tau_{P} \rightarrow \Gamma_P^{tot} $)
of mesons $P =( K^0_L, \, B^0, \, B_s ) $
from Ref. \cite{Tanabashi:2018oca2019}.
For the form factors $f_P$ we use the values \cite{Tanabashi:2018oca}
\begin{eqnarray}
&& 
f_{K^0_L} = f_{K^-} = 155.72 \, MeV, \,\,
f_{B^0} = 190.9 \, MeV, \,\, f_{B_s}= 227.2\, MeV.
\nonumber
\label{eq:fPexp}
\end{eqnarray}
For the mass scale $M_c$ of the chiral four color symmetry
breaking we choose the value $M_c = 10 \,\, \mbox{TeV}$ and for the strong coupling
constant at this mass scale we use the value $\alpha_{st}(10 \,\, \mbox{TeV}) = 0.07134$.

Using (\ref{eq:Br_VLRPlplm})--(\ref{eq:BPl})
and varying the angle $\theta^{L}_{13}$ in
(\ref{eq:betaVLR2B0mues0thete120pi2})--(\ref{eq:betaVLR2Bstaumutaues0thete120pi2})
we have found
the lower mass limit for the chiral gauge leptoquark $V^L$ in the form
\begin{eqnarray}
&&
\hspace{-30mm}
 m_{V^L} \cos{\gamma_L} > 5.68  \,\, \mbox{TeV} \,
\label{eq:mVLlim10}
\end{eqnarray}
where $\cos{\gamma_{L}}$ is defined by the formulas (\ref{eq:cosgammaLR}).
The lightest allowed in (\ref{eq:mVLlim10}) mass $m_{V^L}$ satisfying  
the relation
$m_{V^L} \cos{\gamma_L} = 5.68  \,\, \mbox{TeV} $
in the case of (\ref{eq:solution0}), (\ref{eq:theta120pi2}) is ensured by the value
%
\begin{eqnarray}
&&\hspace{-15mm}
\theta^{L}_{13} = 1.096
\label{eq:theta13}
\end{eqnarray}
of the mixing angle $\theta^{L}_{13}$.

Numerically the mass limit (\ref{eq:mVLlim10}) depends on the relation between
the coupling constants $g^{L}_4$ and $g^{R}_4$.
In particular case of
$\alpha^{L}_{4} = \alpha^{R}_{4} \, ( = 2 \, \alpha_{st}(M_c)= 0.143 ) $
from (\ref{eq:cosgammaLR}) we have
$\cos{\gamma_L}=\cos{\gamma_R}=1/\sqrt 2$
and the eq. (\ref{eq:mVLlim10}) for $g^{L}_4=g^{R}_4$ gives the mass limit
\begin{eqnarray}
&&
\hspace{-30mm}
m_{V^L} > 8.03 \,\, \mbox{TeV} . 
\label{eq:mVLlim20}
\end{eqnarray}
Variation of $\cos{\gamma_L}$ in (\ref{eq:mVLlim10}) according to (\ref{eq:cosgammaLR})
by varying the coupling constants $\alpha^{L,R}_{4}$ in the region ensuring
the validity of the pertubation theory leads to some region for the lower mass limits.
For example with assumption $\alpha^{L}_{4}, \alpha^{R}_{4} < 1$
for $0.08 \lesssim \alpha^{L}_{4} \lesssim 0.5 (1.0) $
the relation (\ref{eq:mVLlim10}) gives
for the values of the lower mass limit the region
\begin{eqnarray}
&&
\hspace{-30mm}
m_{V^L} \gtrsim 6.1 - 15.0 (21.3) \,\, \mbox{TeV}.
\label{eq:mVLlim30}
\end{eqnarray}

The solutions (\ref{eq:solution02}) and (\ref{eq:solution03})
with account of (\ref{eq:betaBsB022to0})
instead of the mass limit (\ref{eq:mVLlim10}) result in the more high mass limits
$m_{V^L} \cos{\gamma_L} > 6.02  \,\, \mbox{TeV}$
and
$m_{V^L} \cos{\gamma_L} > 8.39  \,\, \mbox{TeV}$
respectively.

It is worthy to note that in the case with neglect of fermion mixing ( $K^{L}_2= I )$
the chiral gauge leptoquark $V^L$ contributes only to the decays
$K_L^0 \to \mu e$,  $ B^0 \to \tau e$, $ B_s \to \tau \mu $ and
instead of (\ref{eq:mVLlim10}) we obtain in this case the mass limits
\begin{eqnarray}
&&
\hspace{-30mm}
 m_{V^L} \cos{\gamma_L} > \hspace{3mm} 212.28  \,\, \mbox{TeV} \, , \hspace{3mm}
2.50  \,\, \mbox{TeV} \, , \hspace{3mm}
2.48  \,\, \mbox{TeV} \,
\label{eq:mVLlim1NOmix}
\end{eqnarray}
from the current data on the branching ratios
$Br(K_L^0 \to \mu e)^{exp}$, $Br(B^0 \to \tau e)^{exp}$, $Br(B_s \to \tau \mu )^{exp}$
respectively.
As seen the account of the fermion mixing in leptoquark currents essentially
reduces the mass limit for the chiral gauge leptoquark from
the largest value in (\ref{eq:mVLlim1NOmix}) to the value (\ref{eq:mVLlim10}). 

It is worthy to note also that the mass limits (\ref{eq:mVLlim10}), (\ref{eq:mVLlim20}),
(\ref{eq:mVLlim30}) for the chiral gauge leptoquarks are essentially lower then
corresponding lower mass limit of order of~$90 \,\, \mbox{TeV}$~\cite{Smirnov:2018ske}
for the vector leptoquark.
Such light chiral gauge leptoquarks can manifest themselves in
the leptonic decays of pseudoscalar mesons of type~(\ref{eq:KL0B0Bsdecays})
and in other decays with the lepton flavour violation.

The mass limits (\ref{eq:mVLlim10}), (\ref{eq:mVLlim20}),
(\ref{eq:mVLlim30}) are obtained with neglecting the possible small contributions 
to the diagonal decays $K^0_L \to e^+ e^-, \mu^+ \mu^-$ and 
$ B_s, B^0 \to \mu^+ \mu^- $ from the leptoquark~$V^L$.  
The account of these contributions can result in the small lowering 
of these mass limits. These possible deviations can be approximately estimated 
keeping in mind that the amplitude $M$ and the branching ratio $Br$ 
of the diagonal decay $P \to l^+_i \, l^-_i$ with simultaneous account 
of SM and $V^L$ contributions can be written as 
$ M = M_{SM} +\xi M^{\prime}_{V^L} $ and $Br = Br_{SM} (1 + \xi k + \xi^2 )$ where  
$ \xi $ is a (small) parameter defined as $ \xi^2 = Br_{V^L}/Br_{SM} $ 
with $Br_{V^L}$ defined by expression (\ref{eq:Br_VLRPlplm})  
and $ \xi k $  parametrizes the interference term with $ k \sim 1 $.  
The limiting values of parameter $ \xi $ are defined by the equation      
$ Br_{SM} (1 + \xi k + \xi^2) = Br^{exp}_0 \pm \varDelta Br^{exp}_{\pm} $ 
where $Br^{exp}_0$ and $\varDelta Br^{exp}_{\pm}$ are the central value and 
the experimental errors of the measured branching ratio. These limiting values 
can be found from this equation for each decay using the experimental data 
and the SM predictions for $K^0_L \to e^+ e^-, \mu^+ \mu^-$ and 
$ B_s, B^0 \to \mu^+ \mu^- $ decays.        
The numerical analysis of the part $ Br_{V^L}= \xi^2 Br_{SM} $ 
of the total branching ratio of each of these decays in dependence 
on the leptoquark mass $ m_{V^L} $ leads to the deviation 
of the right side of the relation (\ref{eq:mVLlim10}) by the value     
$ \varDelta m_{V^L} \cos{\gamma_L} \sim - 0.1  \,\, \mbox{TeV} $,  
which induces the corresponding deviations $\varDelta m_{V^L}$ in 
the mass limits~(\ref{eq:mVLlim20}),~(\ref{eq:mVLlim30}). 
In the case of agreement of the future improved measurement of the branching ratio 
of the $ B^0 \to \mu^+ \mu^- $ decay with the SM prediction within the 10\% 
experimental error the deviation $ \varDelta m_{V^L} \cos{\gamma_L} $ 
can be reduced to the value 
$ \varDelta m_{V^L} \cos{\gamma_L} \sim - 0.01  \,\, \mbox{TeV} $.         

We have calculated the contributions to the branching ratios (\ref{eq:Br_VLRPlplm})--
(\ref{eq:BPl})
in the case (\ref{eq:mVRggmVL}) from the chiral gauge leptoquark $V^L$
with the lightest allowed mass $m_{V^L}$ satisfying the relation
$m_{V^L} \cos{\gamma_L} = 5.68  \,\, \mbox{TeV} $
with mixing parameters
(\ref{eq:solution0}),  (\ref{eq:theta120pi2}),  (\ref{eq:theta13}).
These contributions  $Br_{V^L}(P \to l_1 \, l_2)$ are presented in the fourth column
of the Table~\ref{tab:BrVLPl1l2}.
%
%
\begin{table}[h]
\caption{
Contributions 
to the branching ratios
of the decays $P \to l_1 \, l_2$
from the chiral gauge leptoquarks 
with the lightest allowed mass
with account of the general
fermion mixing
a) in the case of $Br_{V^L}(K_L^0 \to l_1 \, l_2)=0$,
$ Br_{V^L}(B_s, B^0 \to \mu^+ \mu^-)=0 $,
$ m_{V^L} \cos{\gamma_L} = 5.68  \,\, \mbox{TeV}$ and
$\theta^{L}_{12} = 0 \,\, (\pi/2) $, 
b) in the symmetrical case of $g^{L}_4=g^{R}_4, \,\, m_{V^L}=m_{V^R} = 8.03 \,\, \mbox{TeV} $
with $\chi ^{R}=\chi ^{L} + \pi $.    
}
\vspace*{3mm}
 \centerline{
\epsfxsize=0.8
\textwidth \epsffile{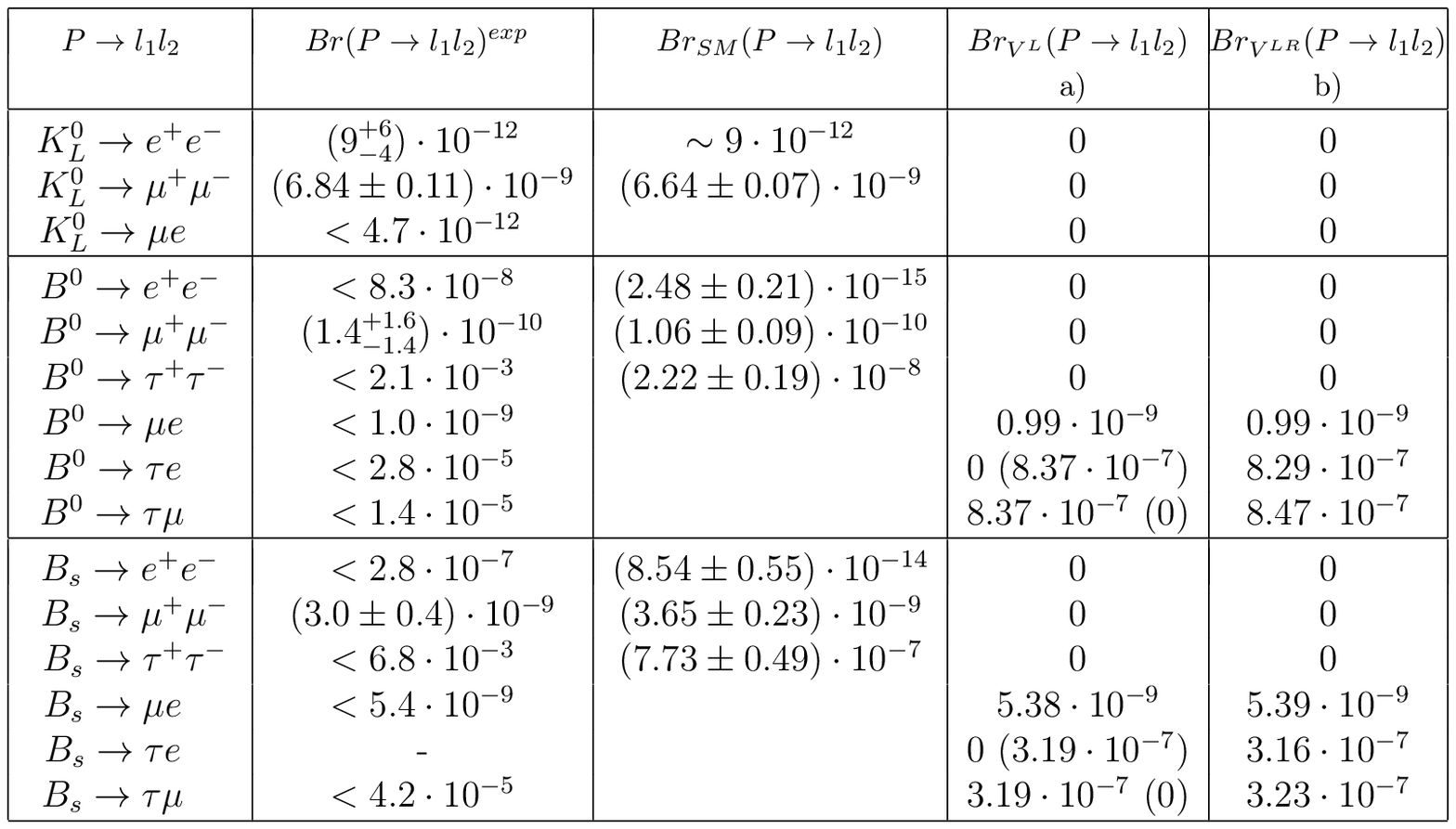}} 
\label{tab:BrVLPl1l2}
\end{table}

As seen the contributions of the chiral gauge leptoquark~$V^L$
to all the diagonal leptonic decays $B_s, B^0 \to l^+_i l^-_i$ of $B_s, B^0$ mesons
(including the decays $ B_s, B^0  \to  e^+ e^-$
and $ B_s, B^0  \to  \tau^+ \tau^-$)
in the case (\ref{eq:betaK0Lijto0}), (\ref{eq:betaBsB022to0})
occur to be equal to zero
whereas such contributions
to $ B_s, B^0 \to \mu e $ decays are close to the corresponding experimental limits.
Just the decays $ B_s, B^0 \to \mu e $ define in this case the lower mass
limit (\ref{eq:mVLlim10})
and the further search for these decays will give the new mass limits
for the chiral gauge leptoquarks.
The analogous contributions to the branching ratios
of the decays $ B_s, B^0  \to  \tau \mu $ (or $ B_s, B^0  \to \tau e $)
are predicted to be of order of $10^{-7}$
and are approximately by order of 2 less than their current
experimental limits (the experimental data on the decays $B_s \to  \tau e$
are still absent) and the search for these decays
will need the more high statistics.

It is interesting to note
that in the case of $m_{V^R} \sim m_{V^L}$
there is the possibility of the mutual cancellation of the contributions
to the branching ratios of the diagonal leptonic decays
$B_s, B^0 \to l^+_i l^-_i$ from the chiral
 gauge leptoquarks~$V^L$ and $V^R $, as it can be seen from the structure
of the mixing factors (\ref{eq:betaVLR2B011})--(\ref{eq:betaVLR2B033}) and
(\ref{eq:betaVLR2Bs11})--(\ref{eq:betaVLR2Bs33}).
In particular, in the symmetrical case with
\begin{eqnarray}
&& \hspace{-15mm}
m_{V^R} = m_{V^L}, \hspace{3mm}  g^{R}_4=g^{L}_4, \hspace{3mm}
(\theta_{ij}, \delta, \varepsilon)^{R} =
(\theta_{ij}, \delta, \varepsilon)^{L}, \hspace{3mm}  \chi^{R}=\chi^{L}+\pi
%
\label{eq:mVRggmVLxx}
\end{eqnarray}
and with account of (\ref{eq:solution0}) we obtain the mass limit
\begin{eqnarray}
&&
\hspace{-30mm}
m_{V^L}=m_{V^R} > 8.03 \,\, \mbox{TeV} , 
\label{eq:mVLVRlim0}
\end{eqnarray}
where the lightest mass 
is ensured in the case (\ref{eq:solution0}) by the angles
\begin{eqnarray}
&&\hspace{-15mm}
\theta^{L}_{12} =\theta^{R}_{12} = 0.78, \hspace{5mm}
\theta^{L}_{13} = \theta^{R}_{13} = 1.096.
\label{eq:theta12theta13LR}
\end{eqnarray}

The contributions  $Br_{V^{LR}}(P\to l_1 l_2)$
to the branching ratios of $P \to l_1 \, l_2$ decays
from the chiral gauge leptoquarks~$V^L$ and $V^R$ with their lightest mass
in (\ref{eq:mVLVRlim0}) and mixing angles
(\ref{eq:solution0}), (\ref{eq:theta12theta13LR})
are presented in the fifth column of the Table~\ref{tab:BrVLPl1l2}.
As seen these contributions to the branching ratios for $ B_s, B^0 \to \mu e $ decays
are also close to the corresponding experimental limits 
as well as those for $ B_s, B^0  \to  \tau e, \tau \mu $  decays
are predicted to be of order of~$10^{-7}$.

The possible contributions of the chiral gauge leptoquarks to the branching ratios
of the decays $P \to l_1 \, l_2$ presented in the Table~\ref{tab:BrVLPl1l2}
are consistent with current data on these decays
and will be useful in designing of the current and future experimental
searches~\cite{Bediaga:2018lhg} for these decays.
The nondiagonal leptonic decays $ B_s, B^0  \to \mu e, \tau e, \tau \mu $ look as
the perspective ones for search for new physics effects such as
the chiral gauge leptoquark contributions to these decays.

In conclusion we resume the results of the work.

The contributions of the chiral gauge leptoquarks $V^L, \, V^R$
induced by the chiral four color symmetry between quarks and leptons
to the branching ratios of $ K_L^0, B^0, B_s \to l_1 \, l_2 $ decays
are calculated and analysed using the general parametrizations
of the fermion mixing matrices in the leptoguark currents.

From the current experimental data on the branching ratios of these decays
the lower mass limit
$ m_{V^L} \cos{\gamma_L} > 5.68  \,\, \mbox{TeV}$
is found (assuming that $ m_{V^L} \ll m_{V^R} $), which gives
$ m_{V^L} > 8.03 \,\, \mbox{TeV} $
for $\alpha^{L}_{4} = \alpha^{R}_{4} $
and
$ m_{V^L} \gtrsim 6.1 - 15.0 (21.3) \,\, \mbox{TeV} $
for $0.08 \lesssim \alpha^{L}_{4} \lesssim 0.5 (1.0) $.

The contributions of the chiral gauge leptoquarks to the branching ratios
of $ K_L^0, B^0, B_s \to l_1 \, l_2 $ decays
are analysed in dependence on the leptoquark masses and the mixing
parameters and the predictions for these contributions to the branching ratios
of $ B^0, B_s \to l_1 \, l_2 $ decays are obtained
in consistency with the current experimental data.

In particular, it is shown that the contributions the chiral gauge leptoquarks
to the diagonal leptonic decays $B_s, B^0 \to l^+_i l^-_i$ can be suppressed
by the appropriate choice of the mixing angles and phases
(in the case of $m_{V^L} \ll m_{V^R}$)
or by the possible mutual cancellation of the contributions of the chiral
gauge leptoquarks to these decays (in the case of $m_{V^L} \sim m_{V^R}$).
In this case with the contributions of the chiral gauge leptoquarks
to the branching ratios of $ B_s, B^0 \to \mu e $ decays
beeng close to the corresponding experimental limits such contributions
to the branching ratios of $ B_s, B^0  \to  \tau e, \tau \mu $  decays
are predicted to be of order of~$10^{-7}$.
The nondiagonal leptonic decays $ B_s, B^0  \to \mu e, \tau e, \tau \mu $ look as
the perspective ones for search for
the chiral gauge leptoquark contributions to these decays.

The estimations of the possible contributions of the chiral gauge leptoquarks
to the branching ratios of the decays $K_L^0, B^0, B_s  \to l_1 \, l_2$
calculated and discussed in this paper with account of the fermion mixing
in leptoquark currents of the general form
and in consistency with the current experimental data
will be useful in designing of the further experimental searches for these decays.

\vspace{3mm} {\bf Acknowledgment}

 The paper is done within
 "YSU Initiative Scientific Researches"
(Project  No.~AAAA-A16-116070610023-3)

\vspace{3mm}


\end{document}